\documentclass[doublecol,final]{epl2} 
\usepackage{amsmath}
\usepackage{amssymb}
\usepackage{cancel}

\title{Novel structure formation of a phase separating colloidal fluid in a ratchet potential}

\author{K. Lichtner\inst{1} \and S. H. L. Klapp\inst{1}}
\shortauthor{K. Lichtner\etal}

\institute{                    
 \inst{1} Institute of Theoretical Physics, Secr. EW~7-1, Technical University Berlin, \\Hardenbergstr. 36, D-10623 Berlin, Germany
}
\pacs{64.75.Xc}{Phase separation and segregation in colloidal systems}
\pacs{47.54.-r}{Pattern selection; pattern formation}
\pacs{64.70.Nd}{Structural transitions in nanoscale materials}

\abstract{
Based on Dynamical Density Functional Theory (DDFT) we investigate a binary mixture of interacting Brownian particles driven over a substrate via a one-dimensional ratchet potential. The particles are modeled as soft spheres where one component carries a classical Heisenberg spin. 
In the absence of a substrate field, the system undergoes a first-order fluid-fluid demixing transition driven by the spin-spin interaction. 
We demonstrate that the interplay between the intrinsic spinodal decomposition and time-dependent external forces leads to 
a novel dynamical instability where stripes \textit{against} the symmetry of the external potential form. 
This structural transition is observed for a broad range of parameters related to the ratchet potential. Moreover, we find intriguing effects for the particle transport.
}

\begin{document}

\maketitle

\section{Introduction}
Understanding the dynamics of particles in complex geometry is an ubiquitary problem throughout non-equilibrium statistical physics with applications in diverse fields such as biology, condensed matter and nanotechnology\cite{RevModPhys.81.387,CPHC:CPHC200800526}. 
Paradigm examples are colloidal particles in periodic optical (or otherwise modulated) potentials \cite{PhysRevLett.96.190601,PhysRevE.75.060101,C0SM01051K}, which display a variety of fascinating transport phenomena including giant diffusion \cite{PhysRevLett.87.010602}, subdiffusive motion \cite{PhysRevLett.98.020602}, and 
ratchet effects, i.e., fluctuating-induced transport
in the absence of a biasing deterministic force \cite{Reimann200257}. 
Indeed, ratchet-driven transport of Brownian (overdamped) particles has been studied in a variety of optical\cite{GrierPRL05,GrierPRE05}, magnetic\cite{PhysRevLett.99.038303,YellenLAB7,PhysRevLett.100.148304,PhysRevLett.105.230602,YellenLAB11,PhysRevLett.109.198304}, and biological systems\cite{bioratchetSCIENCE99,bioratchetELP02,bioratchetNature03}. 
The advantage of studying colloids, which are typically of the size of nano- to micrometer, is that many of these effects can be monitored by real-space experiments (see, e.g., Refs.~\cite{PhysRevE.77.041107,PhysRevLett.106.168104,NJP103017}).\\
In the present letter we study the impact of ratchet potentials on the {\em collective behavior}, specifically the phase separation dynamics, of a colloidal suspension. As a model system we consider 
systems involving {\em magnetic} colloids subject to magnetic ratchet potentials. 
Indeed, recent experimental and theoretical research has shown that magnetic colloidal systems are ideally suited to study transport in complex geometries. Static, magnetic periodic potentials can be created, e.g., by using ferrite garnet films\cite{PhysRevLett.105.230602,PhysRevLett.100.148304,PhysRevLett.99.038303} (where ferromagnetic domains with opposite magnetization direction
are aligned in stripe-like fashion) or by using a periodic arrangement of micromagnets on ``lab-on-a-chip'' devices\cite{YellenLAB7,YellenLAB11,YellenPRE12}. 
An additional time-dependent (oscillating) field can be introduced by combining the static potential with a rotating
magnetic field \cite{YellenLAB7,PhysRevLett.100.148304,PhysRevLett.105.230602,YellenLAB11,PhysRevLett.109.198304}. 
The latter yields a periodic increase (decrease) of the size of domains with parallel (antiparallel) magnetization, which eventually enables transport. From an applicational point of view, a particular attractive feature is that the average velocity of the particles depends on the frequency of the rotating field as well as on the particle geometry \cite{YellenLAB7,PhysRevLett.109.198304}, leading to a
novel approach for particle sorting in mixtures \cite{TiernoSorting09,YellenLAB11}.
Moreover, contrary to electric fields, the properties
of magnetic substrate fields can be changed on the fly (if the frequency is sufficiently small), which is 
an important prerequisite for transport of sensitive objects such as living cells \cite{TiernoPhysChem08,YellenLAB7}.\\
So far, most theoretical studies in this area have been undertaken for {\em single} colloids or systems with negligible interactions. This is in contrast to indications from experiments, which suggest that interaction effects could be important
for the transport in magnetic ratchets \cite{PhysRevLett.109.198304} and the self-assembly of particles into patterns on magnetic lattices \cite{YellenPNAS05}. Only very recently \cite{StraubearXiv}, first theoretical steps have been made to investigate transport of interacting ensembles of magnetic colloids, demonstrating the formation of stable doublets of particles moving over a substrate.\\ 
Here we study collective effects in magnetic ratchet systems arising from an underlying phase separation. To this end we consider a simplified, yet at the same time generic model of a binary mixture of colloids where one species carries a magnetic spin. Experimentally, such systems involve, e.g., ferro-colloids and polymers \cite{PhysRevLett.104.255703}. In our model, the asymmetric interaction potentials yield (in two dimensions) a first-order demixing transition for a broad range of parameters, as we have already shown in earlier studies \cite{lichtner:024502,Lichtner/arXiv}. In the present letter we combine this
interaction potential with a (one-dimensional) ratchet potential coupling to the spins. Our study is based on DDFT \cite{marconi:8032,ArcherSpinDec,espanol:244101}, a generalized diffusion equation where the microscopic interactions enter via the (Helmholtz) free energy. In the last years, DDFT has been successfully applied to a variety of driven systems such as 
colloids in unstable traps \cite{rex08}, sedimenting colloids\cite{royall07} and colloids in washboard potentials with feedback-control\cite{LichtnerEPL,LichtnerPRE86}. The present DDFT results demonstrate that, in combination with a ratchet potential, 
the attractive forces between the magnetic species in the driven mixture lead to a novel instability, that is,
the formation of stripes {\em perpendicular} to the direction of the ratchet potential.

\section{Model}\label{Sec.Theory}
Our model system consists of a binary mixture of Brownian particles confined to a two-dimensional substrate, where one species is magnetic $(m)$ and the other one is non-magnetic $(n)$\cite{lichtner:024502,Lichtner/arXiv} 
(for an experimental counterpart see, e.g., Ref.~\cite{PhysRevLett.104.255703}). 
To model the repulsion between pairs of type $n$-$n$ or $n$-$m$ at positions $\mathbf{r}=(x,z)$ and $\mathbf{r'}=(x',z')$, we employ a Gaussian potential $V_{nn}=V_{nm}=V^\mathrm{core}(|\mathbf{r}-\mathbf{r'}|)=\varepsilon\exp\left[-(\mathbf{r}-\mathbf{r'})^2/\sigma^2\right]$ 
[with $\varepsilon^*=\varepsilon/(k_BT)>0$]. The latter may be considered as a coarse-grained potential for a wide class of ``soft'' (penetrable) colloidal particles (e.g., star polymers\cite{Likos2001267} or dendrimers\cite{GAL06}) with effective (gyration) radius $\sigma$\cite{stillinger:3968}.
The particles from the magnetic species interact via the potential 
$V_{mm}=V^\mathrm{core}(|\mathbf{r}-\mathbf{r'}|)+V^\mathrm{spin}(|\mathbf{r}-\mathbf{r'}|,\omega,\omega')$ where $\omega$ is a set of Euler angles representing the orientation of the unit spin vector $\mathbf{s}$.
For the spin-spin interactions we employ the classical Heisenberg model $V^\mathrm{spin}=J(|\mathbf{r}-\mathbf{r'}|)\mathbf{s}\cdot\mathbf{s'}$ where 
the range dependency is given by Yukawa's potential, that is, $J(|\mathbf{r}-\mathbf{r'}|)=-J\sigma\exp(-|\mathbf{r}-\mathbf{r'}|/\sigma-1)/|\mathbf{r}-\mathbf{r'}|$.
We make the choice $J^*=J/(k_BT)>0$ such that ferromagnetic ordering is favored. We also note that we set $J(|\mathbf{r}-\mathbf{r'}|)=0$ for distances $|\mathbf{r}-\mathbf{r'}|<\sigma$, i.e., we assume that 
at these separations the interaction between two magnetic particles is negligible as compared to the repulsion from the core potentials\cite{lichtner:024502}. 
In fact, for $J^*=0$ the particles are identical and no demixing occurs.\\
To model the magnetic surface fields we use a one-dimensional rocking ratchet potential\cite{Reimann200257}, that is, 
\begin{align}
V_m^\mathrm{ext}(\mathbf{r},t)=&-U \left[\sin\left(0.2 \pi x/\sigma\right)+ 0.25\sin\left(0.4 \pi x/\sigma\right)\right]\nonumber\\ 
&-F\sin(\nu t)x,\label{eq.vext}
\end{align}
where $U$ is the amplitude of the substrate field with broken reflection symmetry and $F$ is the amplitude of an oscillatory driving force with zero mean-force and frequency $\nu$. 
This ratchet potential acts only on the magnetic species. 
In the adiabatic limit $\nu\rightarrow 0$ the net current for a single particle subject to this potential is \emph{always} positive if $F$ exceeds the maximal barrier force, that is, max$(\left\vert\nabla V_m^\mathrm{ext}\right\vert_{F\rightarrow 0})$. 
However, for driving frequencies $\nu>0$ the transport behavior can be rather complex, including current reversal phenomena\cite{bartussek94}. 
\begin{figure}[tpb]
\centering\includegraphics[width=6cm]{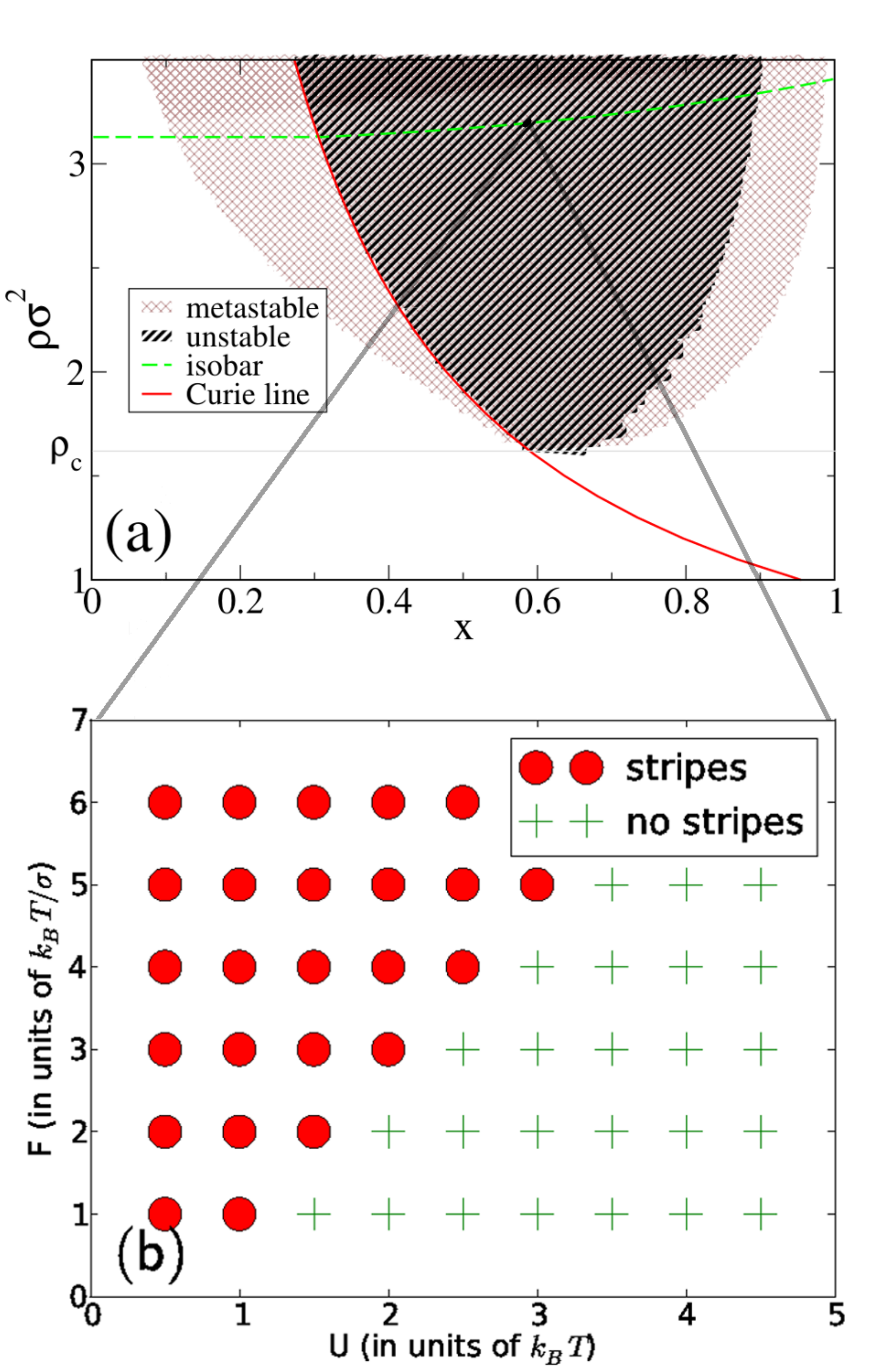}
\caption{(Color online) (a) Bulk phase diagram for the binary $n$-$m$ mixture in the density-concentration plane where $x$ is the concentration of the $m$-particles. The green-dashed line is the $P\sigma^2/(k_BT)=80$ isobar and the Curie line is indicated by the red curve. Unstable and metastable regions are indicated by the shaded regions. (b) Region for stripe formation for the driven system ($\nu=0.1\tau_B^{-1}$) in the $F$-$U$ plane inside the two-phase region. For each point in the phase diagram the density profiles have been calculated for times up to $t=10^4\tau_B$ in order to check for stripe formation. The coupling parameters are $\varepsilon^*=5.0$, $J^*=0.5$.}
\label{fig.phasediag}
\end{figure}
\section{DDFT approach and results}
\begin{figure}[tpb]
\centering\includegraphics[width=6cm]{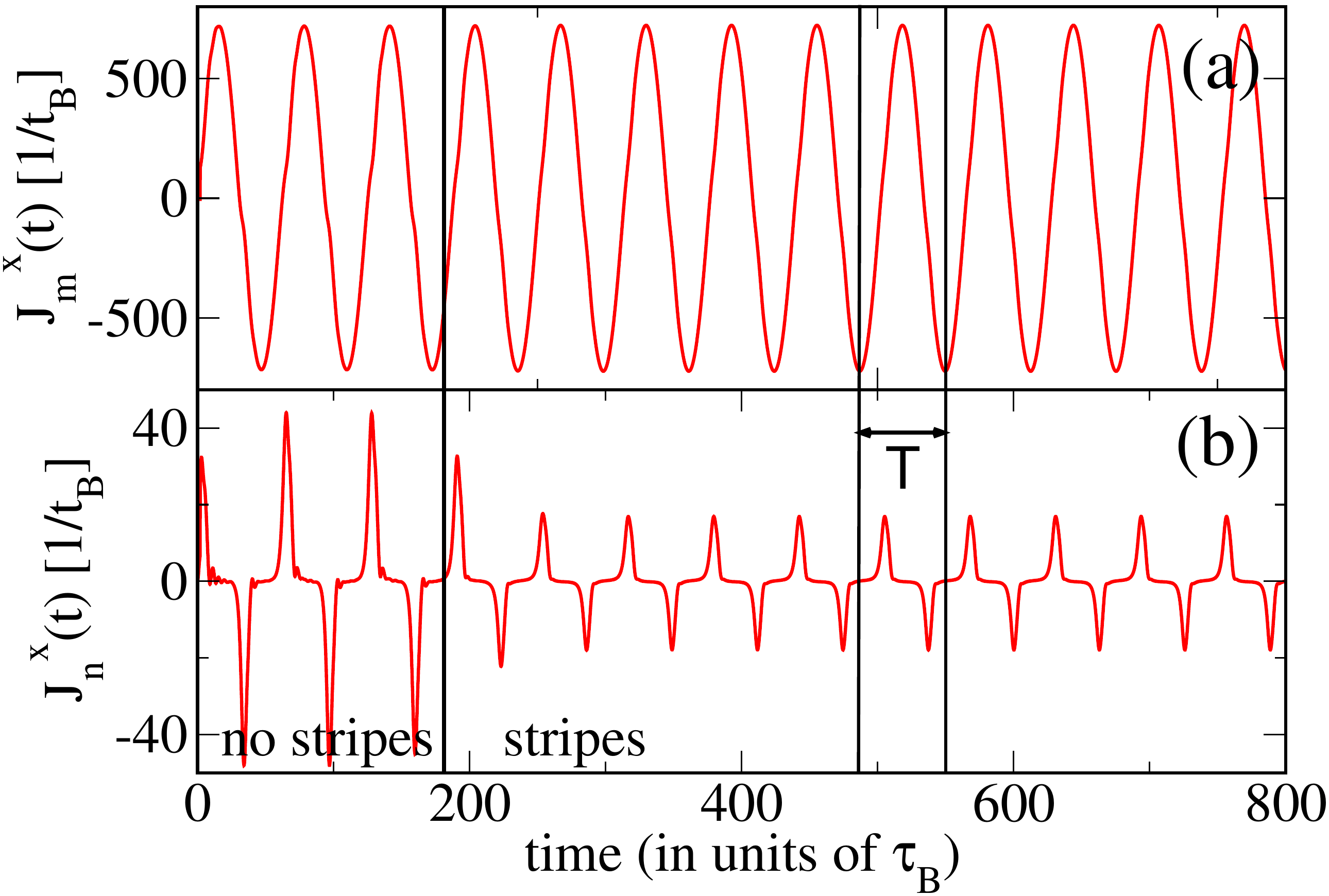}
\caption{(Color online) Averaged particle current along the $x$-direction as a function of time (a) for the magnetic species and (b) for the non-magnetic species. The onset of the longitudinal stripe formation is at $t\simeq 190\tau_B$. The parameters are $U=2k_BT$ and $F=6k_BT/\sigma$.}
\label{fig.timecurrent}
\end{figure}

The non-equilibrium dynamics of the driven system is investigated via a DDFT approach, which in essence is an extension of the classical DFT\cite{evansDFT} (DFT) towards relaxation dynamics. 
The central quantity in DDFT is the time-dependent, one-body density of the (anisotropic) particles, $\rho(\mathbf{r},\omega,t)$.
By construction, the dynamics within DDFT is assumed to be overdamped, i.e., inertial effects are neglected, 
and time-dependent correlations are treated adiabatically. 
Generalizing the DDFT approach towards a binary mixture leads to two coupled integro-differential equations for the density profiles $\rho_\alpha(\mathbf{r},\omega,t)$,\cite{RexLoewen}
\begin{align}
\frac{\partial \rho_\alpha(\mathbf{r},\omega,t)}{\partial t}=&D\nabla \cdot\left[\rho_\alpha(\mathbf{r},\omega,t)\nabla \frac{\delta \mathcal{F}[\{\rho_\alpha(\mathbf{r},\omega,t)\}]}{\delta \rho_\alpha(\mathbf{r},\omega,t)}\right]\label{eq:DDFTmixture}\\
&+D_r\hat{R}\cdot\left[\rho_\alpha(\mathbf{r},\omega,t)\hat{R}\frac{\delta \mathcal{F}[\{\rho_\alpha(\mathbf{r},\omega,t)\}]}{\delta \rho_\alpha(\mathbf{r},\omega,t)}\right].\nonumber
\end{align}
\begin{figure*}[ht]
\centering\hspace*{-1.5cm}
\begin{minipage}[b]{0.23\linewidth}
\centering
\resizebox{55mm}{!}{\includegraphics{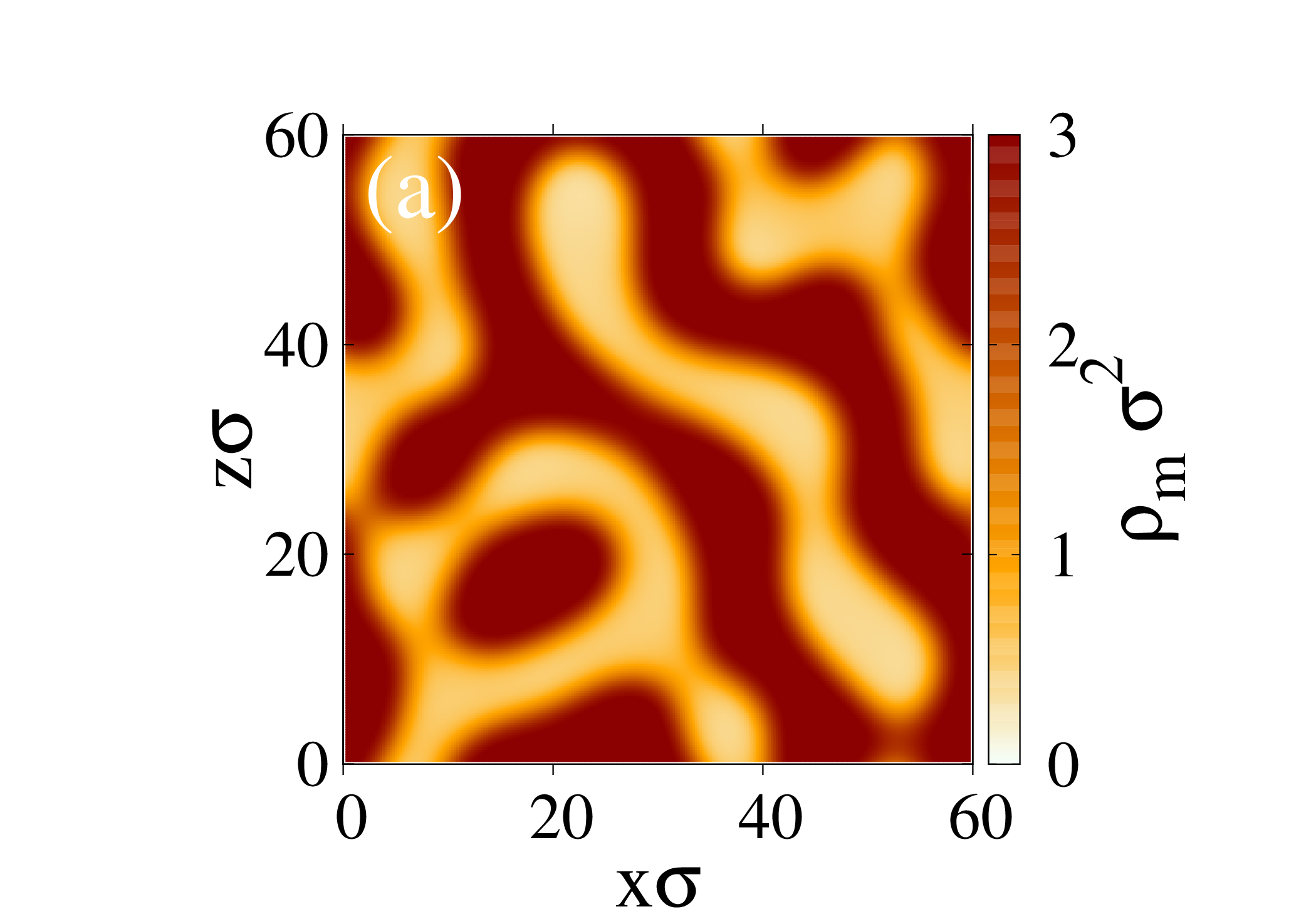}}
\end{minipage}
\hspace{0.2cm}
\begin{minipage}[b]{0.23\linewidth}
\centering
\resizebox{55mm}{!}{\includegraphics{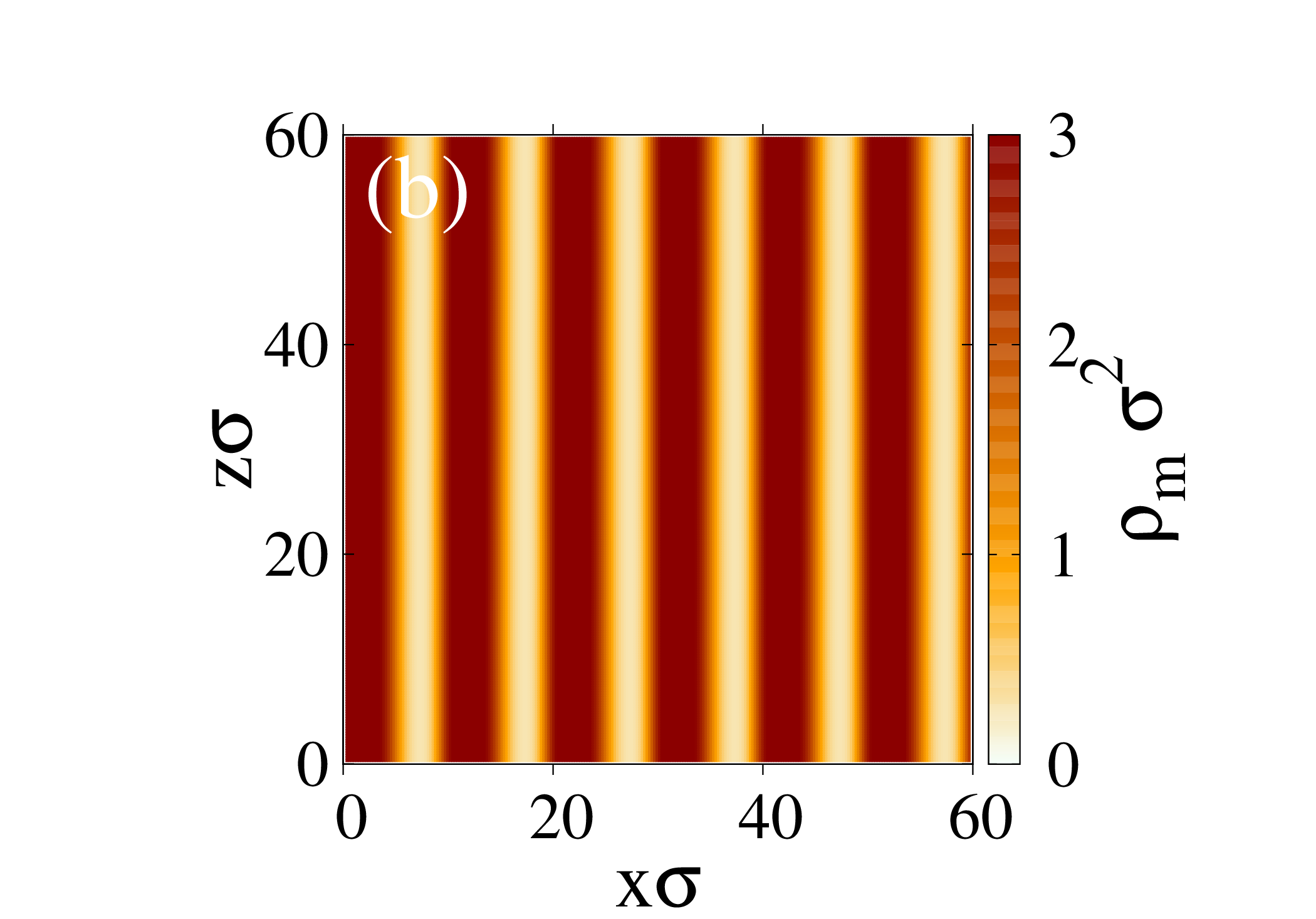}}
\end{minipage}
\hspace{0.2cm}
\begin{minipage}[b]{0.23\linewidth}
\centering
\resizebox{55mm}{!}{\includegraphics{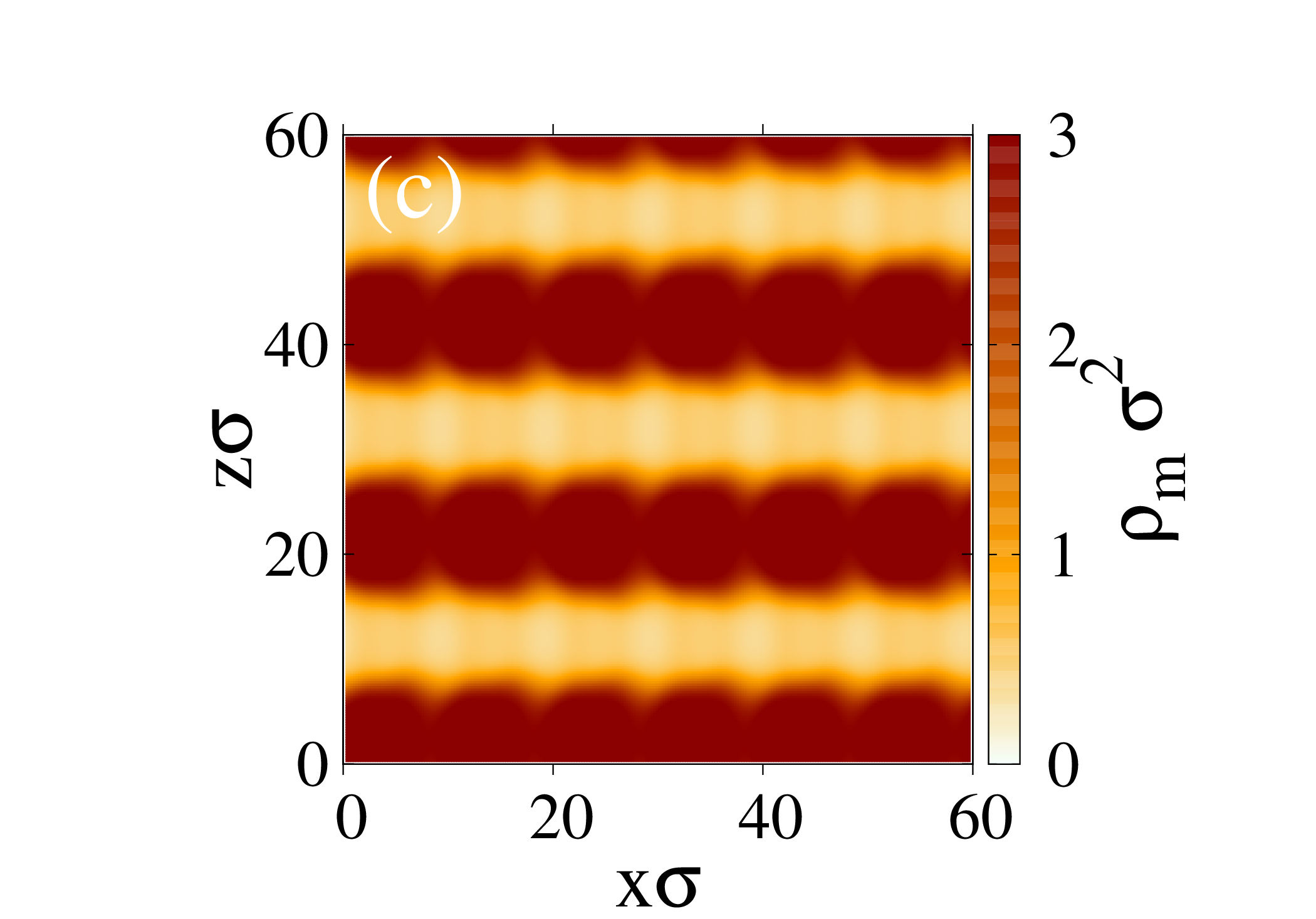}}
\end{minipage}
\hspace{0.1cm}
\begin{minipage}[b]{0.23\linewidth}
\centering
\resizebox{55mm}{!}{\includegraphics{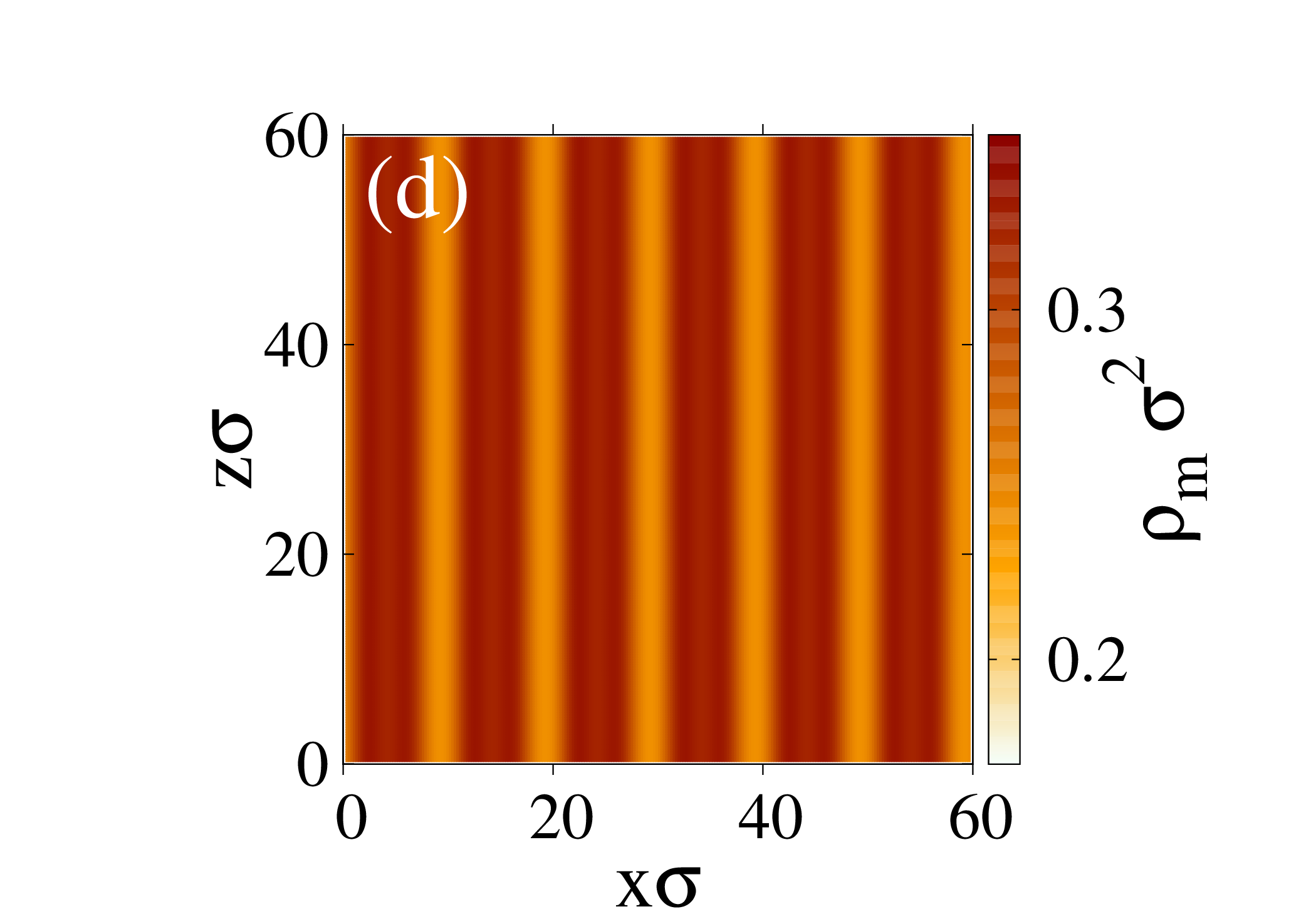}}
\end{minipage}
\caption{(Color online) Typical snapshots of the density profile $\rho_m$ as a function of the position (for simplicity, we only show results for the magnetic species). 
(a) Demixing without external field ($V_m^\mathrm{ext}=0$) at $t=120\tau_B$ for an unstable state ($\rho\sigma^2=3.2$,$x=0.6$) inside the two-phase region [marked point in Fig.~\ref{fig.phasediag}(a)]. 
The impact of the external potential on the demixing is shown for $U=1k_BT$ and $t=1400\tau_B$ (b) for $F=0$ (without drive) and (c) for $F=4k_B T/\sigma$.
(d) Snapshot for the same parameters as in (c) but for a stable state ($\rho\sigma^2=1.5$,$x=0.2$) outside the two-phase region of Fig.~\ref{fig.phasediag}(a).}
\label{fig.rhoplot}
\end{figure*}
where $\hat{R}=\omega\times\nabla_{\omega}$ is the rotation operator and the coefficients $D$ and $D_r$ are the translational and the rotational diffusion constants, respectively. 
The chemical potential in Eq.~(\ref{eq:DDFTmixture}) is determined by the Helmholtz free energy functional $\mathcal{F}$ via the relation $\mu_\alpha=\delta {\cal F}[\{\rho_\alpha\}]/\delta \rho_\alpha$\cite{evansDFT}. Specifically, 
$\mathcal{F}=\mathcal{F}_\mathrm{id}+\mathcal{F}_{\mathrm{ext}}+\mathcal{F}_{\mathrm{ex}}$ where 
$\mathcal{F}_\mathrm{id}=k_BT\sum\limits_{\alpha}\iint d\mathbf{r} d\omega\rho_\alpha(\mathbf{r},\omega,t)
[\ln(\rho_\alpha(\mathbf{r},\omega,t)\Lambda_\alpha^2)-1]$ is the ideal gas part (with $\Lambda_\alpha$ being the thermal de Broglie wavelength of species $\alpha$), 
$\mathcal{F}_\mathrm{ext}=k_BT\sum\limits_{\alpha}\iint d\mathbf{r} d\omega\rho_\alpha(\mathbf{r},\omega,t)V_\alpha^\mathrm{ext}(\mathbf{r},\omega,t)$ is the external
 field contribution (with $V_n^\mathrm{ext}=0$), and $\mathcal{F}_{\mathrm{ex}}$ accounts for particle interactions. 
 For $\mathcal{F}_{\mathrm{ex}}$, we use a mean-field approximation 
$\mathcal{F}_{\mathrm{ex}}=\frac12\sum\limits_{\alpha,\beta}\iiiint d\mathbf{r} d\mathbf{r'} d\omega d\omega'\rho_\alpha(\mathbf{r,\omega},t)
 V_{\alpha\beta}(|\mathbf{r}-\mathbf{r'}|,\omega,\omega')\rho_\beta(\mathbf{r'},\omega',t)$, which is quasi-exact in the high-density limit for ``soft systems'' modeled with a Gaussian core and yields reliable results for the fluid structure even at low and intermediate densities\cite{Likos2001267,PhysRevE.62.7961}. \\
Equation~(\ref{eq:DDFTmixture}) can be simplified drastically by assuming that the magnetic moments relax instantaneously. 
To this end we factorize the one-body density profile into a translational number density part, $\rho_\alpha(\mathbf{r},t)$, and a (normalized) orientational distribution function, $h_\alpha(\mathbf{r},\omega,t)$ and then set
the functional derivative $\delta\mathcal{F}/\delta h_m(\mathbf{r},\omega,t)=0$ for all times $t$ \cite{lichtner:024502}.
This yields a self-consistency relation for the orientational distribution 
$h_m(\mathbf{r},\omega,t)=\exp(\mathbf{B}(\mathbf{r},t)\cdot \mathbf{s}(\omega,t))/\int d\omega \exp(\mathbf{B}(\mathbf{r},t)\cdot \mathbf{s}(\omega,t))$ 
where 
the (self-consistent) effective field is given by 
$\mathbf{B}(\mathbf{r},t)= -\iint d\mathbf{r}' d\omega'\rho_m(\mathbf{r}',t)h_m(\mathbf{r}',\omega',t)J(|\mathbf{r}-\mathbf{r}'|)\mathbf{s'}$ \cite{lichtner:024502}.\\ 
Before we discuss the impact of the full ratchet potential on the dynamics of the system we briefly recall the phase behavior of the ``bulk'' two-dimensional binary mixture with $V_m^\mathrm{ext}=0$ (for details see Ref.~\cite{lichtner:024502}). In Fig.~\ref{fig.phasediag}(a) we show the bulk phase diagram 
for the exemplary case $\varepsilon^*=5.0$ and $J^*=0.5$. The first parameter is well below the ``freezing'' limit such that the system remains fluid at all densities\cite{LLWLoewen2000}.  
We find a demixing phase transition above a critical value $\rho_c$ for the bulk density. This first order phase transition is purely driven by the ferromagnetic interactions as can be seen from the fact that the demixing is coupled to a transition from a paramagnetic phase rich in $n$-particles to a ferromagnetic phase rich in $m$-particles 
[see Fig.~\ref{fig.rhoplot}(a) for a typical snapshot inside the unstable region]. 
The magnetic states are separated from the paramagnetic states by the Curie line [shown as a green-dashed curve in Fig.~\ref{fig.phasediag}(a)]. 
We recall that the coexisting states fulfill the conditions of equal pressure, temperature and chemical potential. 
These states are included in Fig.~\ref{fig.phasediag}(a) as the boundary of the metastable area.\\ 
Now, we consider the system where the magnetic particles are subject to the surface field. 
Without the oscillatory driving force, i.e., setting $F=0$ in $V_m^\mathrm{ext}$ [see Eq.~(\ref{eq.vext})], the external potential becomes static. To illustrate this case we choose a fixed amplitude of the static potential, $U=1k_BT$. 
As can be seen from Fig.~\ref{fig.rhoplot}(b), the density distribution $\rho_m$ is peaked at the minima positions of $V_m^\mathrm{ext}$. 
Thus, the static part of the external potential leads to a symmetry break of the magnetic particle distribution, which is the expected behavior in the purely magnetic system 
(in fact, this behavior is also seen experimentally, see, e.g., Ref.~\cite{YellenPRE12}). 
Due to the repulsive pair interaction, the non-magnetic species is confined to the space where the density $\rho_m$ is small (i.e., at the maxima positions 
of $V_m^\mathrm{ext}$). 
This situation changes when the oscillatory driving force (related to the parameter $F$) is switched on. We consider a fixed 
oscillation frequency $\nu=0.1\tau_B^{-1}$ where the single-particle system displays positive net current \cite{bartussek94}. Increasing $F$ beyond a threshold value $F_c$ (where $F_c(U)$ is a critical value depending on the ratchet amplitude $U$) we observe a shift of the entire density distribution $\rho_m$ (and $\rho_n$) caused by the external force $|\nabla V_m^\mathrm{ext}(\mathbf{r},t)|$. Moreover, after several periods $T$ [e.g., after $50$ periods for $U=1k_BT$ and $F=4k_BT/\sigma$ - see Fig.~\ref{fig.rhoplot}(c)]
we find non-vanishing values for $\rho_m$ for all $x$-positions. In other words, we observe a spontaneous symmetry break of the density distribution indicated by the formation of longitudinal stripes. We interpret this dynamical instability as an interplay between the intrinsic spinodal decomposition and the external magnetic force $\propto|\nabla V_m^\mathrm{ext}(\mathbf{r},t)|$: 
For $F>F_c$, domains of $m$-particles (and $n$-particles through the mutual $m$-$n$ repulsion) are driven over the local barriers of $V_m^\mathrm{ext}$. 
These domains reorganize into stripes since the formation of interconnecting structures is promoted by the bulk system behavior [see Fig.~\ref{fig.rhoplot}(a)]. 
The strict orientation along the $x$-direction reflects the fact that the resulting fluid-fluid interface has to be parallel to the driving force (the study \cite{DzubJPCM02} proofs that curved interfaces are unstable for constant drives). We checked our calculations with respect to system size dependencies by performing trial runs with different box sizes $L$. For the values of $L$ considered, we did not find any impact on the formation of stripes. However, we cannot exclude that the dynamical scaling law for the demixing might be influenced by $L$ (see Ref.~\cite{Lichtner/arXiv} for details). 
Moreover, the positioning of the stripes depends on the initial configuration which is given by the realization of the random white noise applied to the initial bulk profiles \cite{lichtner:024502}.\\
To further highlight the supportive role of the phase separation for stripe formation, we calculate the density distributions $\rho_m$, $\rho_n$ for the same values of $F$ and $U$ but for a bulk set $(\rho,x)$ outside the two-phase region of Fig.~\ref{fig.phasediag}(a). Without the intrinsic phase separation the driving force does not suffice 
for stripe formation [as can be seen in Fig.~\ref{fig.rhoplot}(d)]. 
Another interesting observation is that the intrinsic coarsening process due to spinodal decomposition slows down significantly over the observation time ($\sim 10^4\tau_B$) once the alternating sequence of $m$,$n$ stripes is present\footnote{In the limit $t\rightarrow \infty$ the coarsening process in a binary mixture always leads to entirely demixed macrophases such that the alternating sequence with finite width is, strictly speaking, of transient character.} [see Fig.~\ref{fig.rhoplot}(c)]. 
Thus, the external potential $V_m^\mathrm{ext}(\mathbf{r},t)$ may also be used to suppress spinodal decomposition in one direction contrary to an equivalent system with $V_m^\mathrm{ext}=0$ where no direction for the demixing process is favored (see Fig.~\ref{fig.rhoplot}(a) and Refs.~\cite{lichtner:024502,Lichtner/arXiv}). 
In Fig.~\ref{fig.phasediag}(b) we depict a non-equilibrium state diagram for the stripe formation in the $F$-$U$ plane. 
It is seen that the striped state is separated from the other state by a straight line, i.e., the critical driving force amplitude $F_c$ for stripe formation depends linearly on the ratchet amplitude $U$. In fact, we find 
$F_c\sigma =2\mathrm{max}(\left.V_m^\mathrm{ext}\right\vert_{F\rightarrow 0})-\Delta(J,\varepsilon)$ where $\mathrm{max}(\left.V_m^\mathrm{ext}\right\vert_{F\rightarrow 0})\simeq 1.1U$ is the energy barrier of the static part of $V_m^\mathrm{ext}$ and $\Delta(J,\varepsilon)$ is a constant 
depending on the internal interactions of the system (e.g., $\Delta\simeq 1.5k_BT$ for $J^*=0.5$, $\varepsilon^*=5$). 
Since the stripe formation is induced by the ferromagnetic coupling, we expect $F_c(U)$ to decrease if $J$ is being increased. 
We have not observed any hysteresis effects in the $F$-$U$ plane.\\
We now explore the impact of the observed dynamical instability on the transport properties of the system. To this end we calculate the density current given by the divergence term in Eq.~(\ref{eq:DDFTmixture}), that is, 
$\mathbf{J}_\alpha=-D\rho_\alpha\nabla\left(\delta {\cal F}[\{\rho_\alpha\}]/\delta \rho_\alpha\right)$. 
Specifically, we are interested in the particle currents along the x-direction that we obtain by averaging over all $z$-positions, that is,
\begin{align}
J_\alpha^x(t)=\frac{1}{N_z}\sum\limits_{i=1}^{N_z}\int\limits_{0}^{L} dx J^x_\alpha(x,z_i,t), \;\;\;\alpha=\{m,n\},
\label{eq.xcurrenttime}
\end{align}
where $N_z$ is the number of discretization points in $z$-direction.
\begin{figure}[tpb]
\centering\includegraphics[width=6cm]{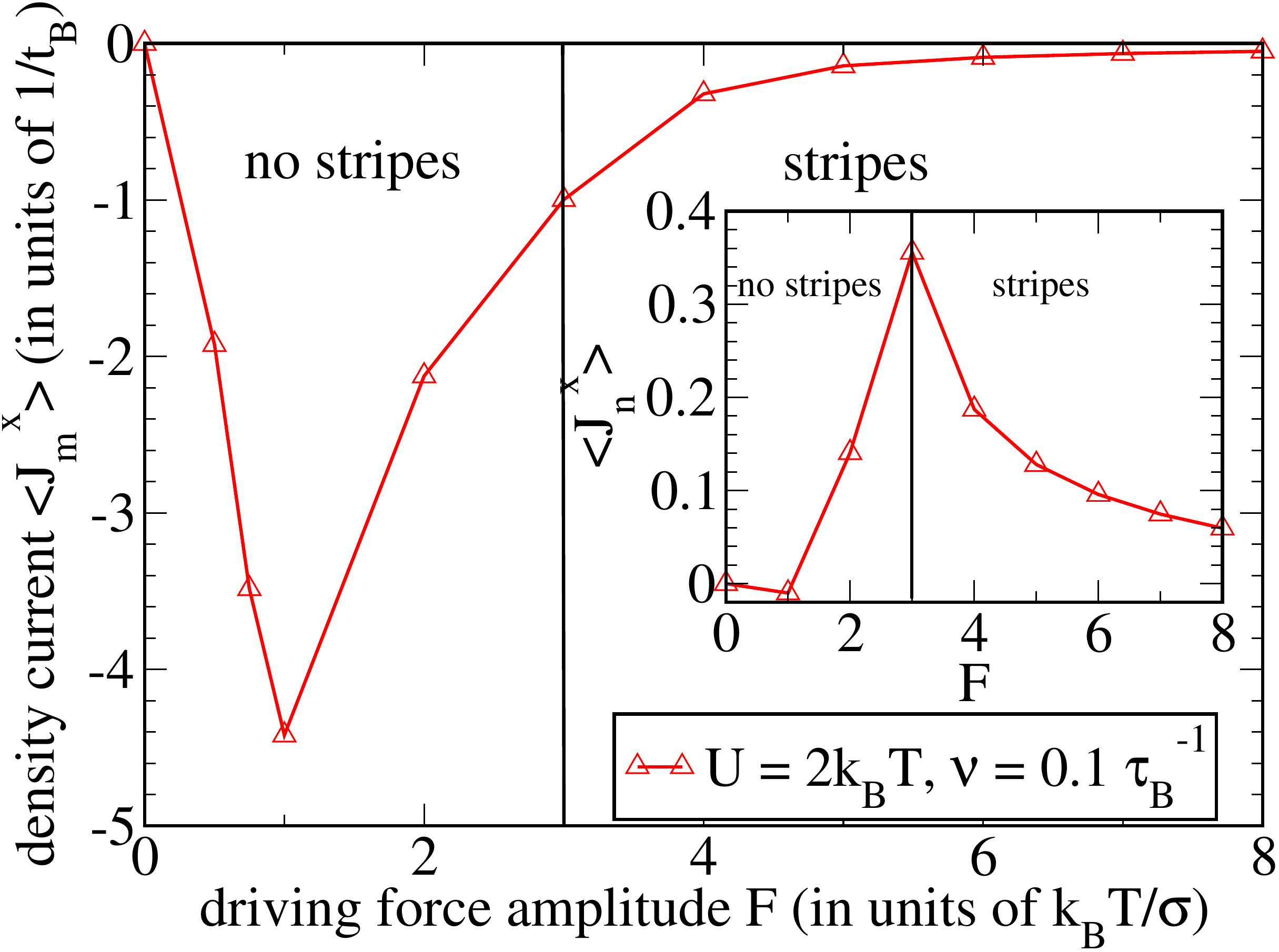}
\caption{(Color online) Net particle current $\langle J_m^x\rangle$ for the magnetic species as a function of $F$. The inset shows the same curve for the non-magnetic species. The critical driving force $F_c$ is indicated by the vertical lines.}
\label{fig.stromrhom}
\end{figure}
In Fig.~\ref{fig.stromrhom} we show results for the resulting net current that we derive from Eq.~(\ref{eq.xcurrenttime}) by time-averaging over one period $T$, that is,
$\langle J_\alpha^x\rangle=1/T\int_{\tilde{t}}^{\tilde{t}+T} dt J^x_\alpha(t)$ 
where $\tilde{t}$ is a time after the initial transient period (i.e., after the onset of the stripe formation). 
For a fixed value of $U=2k_BT$ we follow a path through Fig.~\ref{fig.phasediag}(b) by varying $F$.\\
For small values of $F$ we observe a 
net transport of $m$-particles in the {\em backward} direction, as can be seen from the negative values of $\langle J^x_m\rangle$. We note that 
this is in contrast to the behavior of a single particle
in the ratchet potential, which would display {\em positive} current at the parameters considered. 
We suspect that this difference is due to an interaction effect, specifically the repulsion between magnetic and non-magnetic particles which sit at the potential maxima [see Fig.~\ref{fig.rhoplot}(b)] and thus effectively increase the potential height at the parameters considered. 
We also note that the actual values of $\langle J^x_m\rangle$ strongly depend on the parameter $U$. For the value $U=2k_BT$ considered in Fig.~\ref{fig.stromrhom}, the magnitude of the negative current becomes maximal for $F=1k_BT/\sigma$. 
As $F$ is being increased further towards the threshold value $F_c\vert_{U=2}\simeq 3k_BT/\sigma$ we find that the ratchet effect is decreasing again. Hence, the stripe formation 
effectively decreases the reflection asymmetry of the potential. 
In the inset we show the result for the non-magnetic species. Similarly to the magnetic species, we find that the transport of non-magnetic particles is suppressed within the stripe forming region of Fig.~\ref{fig.phasediag}(b). On the other hand, for the non-striped states we find a finite net current $\langle J^x_n\rangle$ 
in the direction \textit{opposite} to the current of the magnetic particles. 
This is based on the fact that the non-magnetic species ``sees'' effectively a mirrored potential with inverse reflection asymmetry due to the repulsive pair interaction with the other species. 
Moreover, $\langle J^x_n\rangle$ exhibits a peak as a function of $F$ in the region where the system transitions into the striped state. From the density profiles we conclude that this behavior is linked to the repositioning of the non-magnetic species 
at the onset of the stripe formation: By further increasing $F$ the particle separation into distinct longitudinal stripes reduces the total number of collisions between different species resulting in a decreased current $|\langle J^x_n\rangle|$. 
To show the dynamical behavior as the system undergoes the structural transition we exemplarily calculate with Eq.~(\ref{eq.xcurrenttime}) the \textit{instantaneous} current for a data point of Fig.~\ref{fig.phasediag}(b) within the stripe forming region. 
As can be seen from Fig.~\ref{fig.timecurrent}(a) the current for the $m$-particles oscillates between $\pm 720 \tau_B^{-1}$ with period $T\simeq 62.8\tau_B$ indicating a back-and-forth rocking motion. 
Due to the $n$-$m$ repulsion the same behavior is seen for the other species [see Fig.~\ref{fig.timecurrent}(b)] - 
the large quantitative difference between $J^x_n$ and $J^x_m$ reflects the fact that $V_m^\mathrm{ext}$ only couples to the $m$-particles. 
Furthermore, we observe a sudden drop in the amplitude of the $J_n^x$-oscillations at $3T\simeq 190\tau_B$ [see Fig.~\ref{fig.timecurrent}(b)]. 
From inspecting the density snapshots we find that this current drop is linked to the onset of the global stripe formation. 
Thus, the behavior of the $n$-particles serves again as a true indicator for signaling the transition into the striped state (cf. Fig.~\ref{fig.stromrhom}).

\section{Concluding remarks}
In this letter, we have demonstrated that the (non-equilibrium) pattern formation in a phase separating colloidal mixture can be efficiently manipulated by applying external time-dependent surface fields (represented here by the one-body potential $V_m^\mathrm{ext}$). 
For $V_m^\mathrm{ext}=0$, the system exhibits a first-order demixing transition accompanied by spinodal decomposition. 
This occurs in the region of the bulk phase diagram where the system is unstable against harmonic density perturbations with certain wave numbers $\left\vert \mathbf{k}\right\vert$\cite{Lichtner/arXiv}. 
However, there is no favored direction for the demixing process. This symmetry is already broken if we switch on a static external potential ($F\rightarrow 0$) acting on the magnetic species. 
Moreover, for the driven system ($F> 0$) we demonstrated that the \emph{direction} of the symmetry break can be changed if the condition $F\sigma\geq 2\mathrm{max}(\left.V_m^\mathrm{ext}\right\vert_{F\rightarrow 0})-\Delta$ is met where the first term on the right hand side is the energy barrier of 
the static part of the external potential and $\Delta=\Delta(J,\varepsilon)$ 
is a constant depending on the internal interactions of the system. Moreover, the latter structural transition (that we call \emph{stripe formation}) suppresses the transport of particles along the $x$-direction that is otherwise observed due to the magnetic ratchet effect. 
Indeed, the transport behavior turns out to be non-trivial due to the collective effects but remains tunable by the parameters $U$ and $F$. 
We stress that the stripe formation is a combinatory effect of the underlying demixing transition and the (one-body) interactions with the external surface fields. Indeed, without the demixing we cannot find any transition for the values $F/U$ considered here. 
Moreover, we note that the stripe formation is quite robust against changes of the ratchet amplitude $U$, driving amplitude $F$ (with $F/U$ being constant), as well as the total density $\rho$ and concentration $x$ (inside the two-phase region). Indeed, for closed systems (i.e., without particle exchange), the parameter $x$ seems to be a suitable parameter for controlling the size of the globally forming stripes. We also observed robustness of the stripe formation against the frequency $\nu$ (for the range $0.05$-$1\tau_B^{-1}$). 
For higher frequencies (e.g., $\nu\gtrsim 5\tau_B^{-1}$ for $F\sigma/U=1$) current reversal phenomena have been reported in the single-particle limit \cite{bartussek94}. 
Clearly, a follow-up study covering the full frequency range would be interesting for the present system. 
We note that pattern formation in phase separating systems may also be controllable by applying other types of external forcing. 
Indeed, a recent theoretical study for the Langmuir-Blodgett transfer reveals that the self-organization of the particles may be manipulated by changing, both, the substrate properties and the transfer velocity\cite{Wilczek2013}. 
In another study \cite{Weith2009} the effects of a spatially periodic forcing traveling with constant velocity $v$ on the coarsening has been studied based on a Cahn-Hilliard approach. 
However, we stress none of these studies \cite{Wilczek2013,Weith2009} report periodic solutions against the symmetry of the external field as we have found here.\\
An experimental realization of the present results seems possible, e.g., by employing polystyrene (paramagnetic) particles of size $\sigma=1.4\mu$m on a 2D surface with parallel magnetic stripes created by a ferrite garnet film with spatial periodicity $\lambda=6.9\mu$m$\simeq 4.9\sigma$\cite{PhysRevLett.105.230602}. 
The translation along the horizontal direction is technically feasible by superimposing an external, rotating magnetic field where the target velocity is tunable via the frequency\cite{YellenLAB7,PhysRevLett.105.230602}. 
Typical travel distances within a period $2\pi/\nu$ in Ref.~\cite{PhysRevLett.109.198304} are $0 -16 \sigma$ compared to $0-14\sigma$ in the present study, suggesting that our parameters ($\lambda=10\sigma$,$0 \leq F\sigma/U\leq 10$, $\nu=0.1\tau_B^{-1}$) are not unrealistic. 
Furthermore, we note that all quantities in the present colloidal system are,
in principle, accessible by experiments \cite{PhysRevE.77.041107,PhysRevLett.106.168104,NJP103017}.
We therefore hope that our
results will stimulate future experiments. 
Clearly, from the theoretical side it would be desirable to extend the present study to colloids with true dipolar (instead of Heisenberg) interactions, which enable directed self-assembly. 
Another interesting idea is to supplement the modulated potential by a feedback control force, i.e., a force depending on the state of the system. 
In fact, a first experimental realization of feedback-controlled currents in flashing ratchets already exists \cite{PRL08Linke}. 
This could be a promising route for the development of novel particle assemblers on the nano to micro-scale.

\acknowledgements
We gratefully acknowledge financial support via the Collaborative Research Center (SFB) 910.

\bibliographystyle{eplbib} 

\providecommand{\noopsort}[1]{}\providecommand{\singleletter}[1]{#1}%

\end{document}